# COMPLEX SYSTEMS ANALYSIS OF CELL CYCLING MODELS IN CARCINOGENESIS: II.

## Cell Genome and Interactome, Non-random, Nonlinear Dynamic Models in Łukasiewicz Logic Algebras-- Neoplastic Transformation Models in Łukasiewicz-Topos with Pseudo-Markov Chain Processes representing Progression Stages during Carcinogenesis and Cancer Therapy

I. C. Baianu
University of Illinois at Urbana,
Urbana, IL 61801,
USA

email: **ibaianu@uiuc.edu**

ABSTRACT

**Carcinogenesis is a complex process that involves dynamically inter-connected modular sub-networks that evolve under the influence of micro-environmental, as well as, in many cases, cancer therapy-induced perturbations, in non-random, pseudo-Markov chain processes. An appropriate n-stage model of carcinogenesis involves therefore n-valued Logic treatments of such processes, nonlinear dynamic transformations of complex functional genomes and cell interactomes. Lukasiewicz Algebraic Logic models of genetic networks and signaling pathways in cells are formulated in terms of nonlinear dynamic systems with n-state components that allow for the generalization of previous, Boolean or "fuzzy", logic models of genetic activities in vivo. Such models are then applied to cell transformations during carcinogenesis based on very extensive genomic transcription and translation data from the CGAP databases supported by NCI. Inter-related signaling pathways include very large numbers of different biomolecules, such as proteins, in the intercellular, membrane, cytosolic, nuclear and nucleolar compartments. One such family of pathways contains the cell cyclins. Cyclins are proteins that link several critical pro-apoptotic and other cell cycling/division components, including the tumor suppressor gene TP53 and its product, the Thomsen-Friedenreich antigen (T antigen), Rb, mdm2, c-Myc, p21, p27, Bax, Bad and Bcl-2, which all play major roles in carcinogenesis of many cancers. A categorical and Lukasiewicz-Topos (LT) framework for Lukasiewicz Algebraic Logic models of nonlinear dynamics in complex functional genomes and cell interactomes. An algebraic formulation of varying 'next-state functions' is extended to a Łukasiewicz Topos with an n-valued Łukasiewicz Algebraic Logics subobject classifier description that represents non-random and nonlinear network activities as well as their transformations in developmental processes and carcinogeness. Specific models for different types of cancer are then derived from representations of the dynamic state-space of LT non-random, pseudo-Markov chain process, network models in terms of cDNA and proteomic, high throughput analyses by ultra-sensitive techniques. This novel theoretical analysis is based on extensive CGAP genomic data for human tumors , as well as recently**

**published studies of cyclin signaling, with special emphasis placed on the roles of cyclins D1 and E. Several such specific models suggest novel clinical trials and rational therapies of cancer through re-establishment of cell cycling inhibition in stage III cancers.**

*1. Introduction*.

The calculus of predicates of formal Hilbert Logic was applied by Nicolas Rashevsky (1965) to generate organismic set theories based on relational biology. Lőfgren (1968) subsequently introduced a quite different, more general logical approach than that of Rashevsky's predicate calculus to the problem of self-reproduction. An attempt to provide a characterization of genetic activities in terms of n-valued logics and generalized biodynamic state-spaces is introduced. For operational reasons the model is directly formulated in an algebraic form by means of Łukasiewicz algebras. The Łukasiewicz algebras were introduced by Moisil (1940) as algebraic models of *n*-valued logics, and further improvements were made by utilizing categorical constructions of Łukasiewicz Logic algebras (Georgescu and Vraciu, 1970).

Had the structural genes presented an "all-or-none" type of response to the action of regulatory genes, a very simple neural nets might have some partial dynamical analogy to a correspondingly small genetic network. Then, both types of net would be only two distinct realizations of a net which is built up of two-factor elements (Rosen, 1970). This would allow for a detailed dynamica1 analysis of their action (Rosen, 1970). However, the case which we consider first is the one in which the activity of the genes is *not* necessarily of the "all-or-none" type. Nevertheless, the representation of elements of a net (in our case these are genes, operons, or groups of genes), as black boxes is convenient for formal reasons, and will be maintained in the sequel (see Figure 1).

**2. Nonlinear Dynamics in Non-Random Genetic Network Models in Łukasiewicz Logic Algebras**.

Jacob and Monod (1961) have shown, that in *E. Coli* the "regulator gene" and three "structural genes" concerned with lactose metabolism lie near one another in the same region of the chromosome. Another special region near one of the structural genes has the capacity of responding to the regulator gene, and it is called the "operator gene". The three structural genes are under the control of the same operator and the entire aggregate of genes represents a functional unit or "operon". The presence of this "clustering" of genes seems to be doubtful in the case of higher organisms, and therefore, more complex networks of genes and genetic network modules are now being extensively studied in conjunction with genomic and proteomic data analysis for medical-oriented purposes, such as individualized cancer therapy development. Carcinogenesis is a complex process that involves dynamically inter-connected modular sub-networks that evolve under the influence of micro-environmental, as well as, in many cases, cancer therapy-induced perturbations, in non-random, pseudo-Markov chain processes. An appropriate n-stage model of carcinogenesis involves therefore n-valued Logic treatments of such processes, nonlinear dynamic transformations of complex functional genomes and cell interactomes. Lukasiewicz Algebraic Logic models of genetic networks and signaling pathways in cells are formulated in terms of nonlinear dynamic systems with n-state components that allow for the generalization of previous, Boolean or "fuzzy", logic models of genetic activities in vivo. Such models are then applied to cell transformations during carcinogenesis based on very extensive genomic transcription and translation data from the CGAP databases supported by NCI. Inter-related signaling pathways include very large numbers of different biomolecules, such as proteins, in the intercellular, membrane, cytosolic, nuclear and nucleolar compartments. One such family of pathways contains the cell cyclins. A detailed model of cell cyclins in carcinogenesis was recently presented (Baianu and Prisecaru, 2004; arXiv.q-bio.OT/0406046 *Preprint*). Thus, it would be natural to term any assembly, or aggregate, of interacting genes as a *genetic network*, without considering the 'clustering' of genes as a necessary condition for all biological organisms. Had the structural genes presented an "all-or-none" type of response to the action of regulatory genes, the neural nets might be considered to be dynamically analogous to the corresponding genetic networks, especially since the former also have coupled , intra-neuronal signaling pathways resembling-but distinct- from those of other types of cells in higher organisms. In a broad sense, both types of network could be considered as two distinct realizations of a network which is built up of two-factor elements (Rosen, 1970). This allows for a detailed dynamica1 analysis of their action (Rosen, 1970). However, the case that was considered first as being the more suitable alternative (Baianu, 1977) is the one in which the activities of the genes are *not* necessarily of the "all-or-none" type. Nevertheless, the representation of elements of a net (in our case these are genes, operons, or groups of genes), as black boxes is convenient, and is here retained to keep the presentation both simple and intuitive (see Figure 1). Previously, the assumption was made (Baianu,1977) that certain genetic activities have **_n_** levels of intensity, and this assumption is justified both by the existence of epigenetic controls, as well as by the coupling of the genome to the rest of the cell through specific signaling pathways that are involved in the modulation of both translation and transcription control processes. This model is a description of genetic activities in terms of *n*-valued Łukasiewicz logics. For operational reasons the model is directly formulated in an algebraic form by means of Łukasiewicz Logic algebras. The formalization of genetic networks that was introduced previously (Baianu, 1977) in terms of Łukasiewicz Logic, and the appropriate definitions are here briefly recalled in order to maintain a self-contained presentation.

The genetic network presented in Figure 1 is a discriminating network (Rosen, 1970). Let us consider the system in Figure 1b and apply to it a simple system formalization that converts the system into logical expressions with n-values. The level (chemical concentration) of *P*1. is zero when the operon *A* is inactive, and it will take some definite non-zero values on levels '1', '2', and '(*n*-1)*'*, otherwise. The first of A is obtained for a threshold value $\delta$ of P2-that corresponds to a certain level of 'j' *of B*. Similarly', the other corresponding thresholds for levels 1,2,3,... and'(n-1)' are, respectively, $u_1.^A$:,. $U_2.^A$ $u_2.^A$ $u_{n-1}.^A$. The thresholds are indicated inside the black boxes, in a sequential order, as shown in Figure 2. Thus, if *A* is inactive (that is, on the zero level), then *B* will be active on the *k* level which is characterized by certain concentration of $P_2$. Symbolically, we write: A(t;0) .≡ . B(t+$\delta$ ,k), where t denotes time and $\delta$ is the 'time lag' or delay after which the inactivity of A is reflected in to the activity of B, on the *k* level. Similarly, one has:

A(t' + ε,n-1).≡. B(t';0).

The levels of A and B, as well as the time lags $\delta$ and ε, need not be the same, More complicated situations arise when there are many concomitant actions on the same gene. These situations are analogous to a neuron with alterable synapses. Such complex situations could arise through interactions which belong to distinct metabolic pathways.

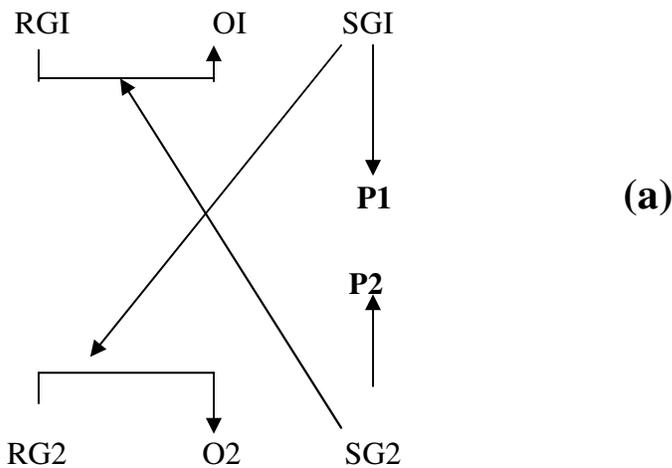

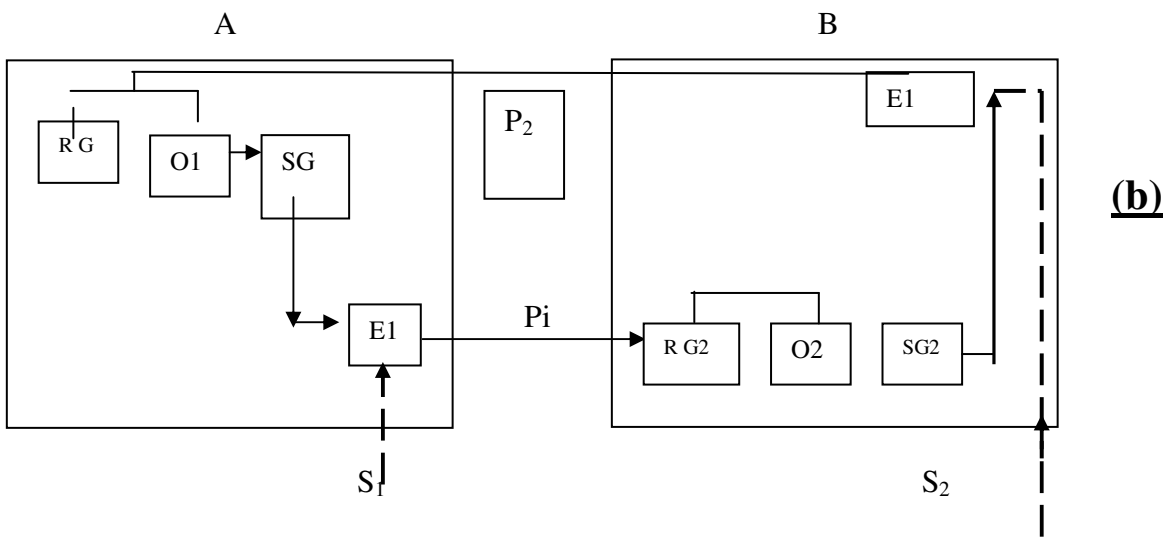

**Figure 1**. The simplest control unit in genetic net and its corresponding black-box images.

The levels of A and B, as well as the time lags δ and ε, need not be the same, More complicated situations arise when there are many concomitant actions on the same gene. These situations are analogous to a neuron with alterable synapses. Such complex situations could arise through interactions which belong to distinct metabolic pathways. In order to be able to deal with any particular situation of this type one needs the symbols of n-valued logics. Re-label the last (n-1) level of a gene by 1. An intermediate level of the same gene should be then relabeled by a lower case letter, x or y. The zero level will be labeled by '0', as before. Assume that the levels of all other genes can be represented by intermediate levels. (It is only a convenient convention and it does not impose any further restriction on the number of situations which could arise). With all assertions of the type "gene *A* is active on the *i*-th level and gene B is active on the *j*-th level" one can form a distributive lattice, L. The composition laws for the lattice will be denoted by ∪ and ∩.

The symbol ∪ stand for the logical non-exclusive 'or', and ∩ will stand for the logical conjunction symbol '**and**'.

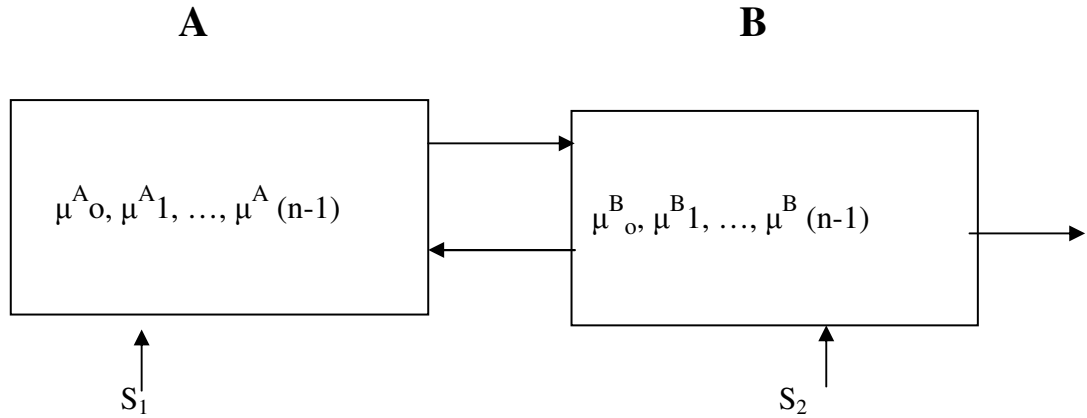

**Figure 2**. Black-boxes with *n* levels of activity.

The levels of A and B, as well as the time lags δ and ε, need not be the same, More complicated situations arise when there are many concomitant actions on the same gene. These situations are analogous to a neuron with alterable synapses. Such complex situations could arise through interactions which belong to distinct metabolic pathways. In order to be able to deal with any particular situation of this type one needs the symbols of n-valued logics. Re-label the last (n-1) level of a gene by 1. An intermediate level of the same gene should be then relabeled by a lower case letter, x or y. The zero level will be labeled by '0', as before. Assume that the levels of all other genes can be represented by intermediate levels. (It is only a convenient convention and it does not impose any further restriction on the number of situations which could arise). With all assertions of the type "gene *A* is active on the *i*-th level and gene B is active on the *j*-th level" one can form a distributive lattice, L. The composition laws for the lattice will be denoted by ∪ and ∩. The symbol ∪ will stand for the logical non-exclusive 'or', and ∩ will stand for the logical conjunction 'and'.

Another symbol "⊂" allows for the ordering of the levels and is the canonical ordering of the lattice. Then, one is able to give a symbolic characterization of the dynamics of a gene of the not with respect to each level i. This is achieved by means of the maps $\delta_t: L \rightarrow L$ *and* $N: L \rightarrow L$, (with *N* being the *negation*). The necessary logical restrictions on the actions of these maps lead to **an n-valued Łukasiewicz algebra**.

(I) There is a map N: L →L, so that N(N(X))= X, N(X ∪Y) = N(X) ∩N(Y) and

N(X∩Y) = N(X) ∪N(Y), for any X, Y ∈ L.

(II) there are (n-1) maps $\delta i$: L→L which have the following properties

(a) $\delta i(0) = 0$, $\delta i(1) = 1$, for any i=1,2,….n-1;

(b) $\delta i(X \cup Y) = \delta(X) \cup \delta i(Y)$, $\delta i(X \cap Y) = \delta i(X) \cap \delta i(Y)$, for any X, Y ∈ L, and i=1,2,…, n-1;

(c) $\delta i(X) \cup N(\delta i(X)) = 1$, $\delta i(X) \cap N(\delta i(X)) = 0$, for any X ∈ L;

(d) $\delta i(X) \subset \delta 2(X) \subset … \subset \delta n-1(X)$, for any X ∈ L;

(e) $\delta h * \delta k = \delta k$ for h, k =1, …, n-1;

(f) I f $\delta i(X) = \delta i(Y)$ for any i=1,2,…, n-1, then X=Y;

(g) $\delta i (N(X)) = N(\delta j(X))$, for i+j =n.

(Georgescu and Vraciu, 1970).

The first axiom states that the double negation has no effect on any assertion concerning any level, and that a simple negation changes the disjunction into conjunction and conversely. The second axiom presets in the fact ten sub cases which are summarized in equations (a) –(g). Sub case (IIa) states that the dynamics of the genetic net is such that it maintains the genes structurally unchanged. It does not allow for mutations which would alter the lowest and 'the highest levels of activities if the genetic net, and which would, in fact, change the whole net. Thus, maps $\delta$: L→L are chosen to represent the dynamical behavior of the genetic nets in the absence of mutations.

Equation (IIb) shows that the maps $\delta$ maintain the type of conjunction and disjunction.

Equations (IIc) are chosen to represent assertions of the following type.

<the sentence "a gene is active on the *i-th* level *or* it is inactive on the same level" is true), and <the sentence "a gene is inactive on the i-th level *and* it is inactive on the same level" is always false>.

Equation (IId) actually defines the actions of maps $\delta t$. Thus, "I is chosen to represent a change from a certain level to a level as low as possible, just above the zero level of L. $\delta 2$ carries a certain level x in assertion X just above the same level in $\delta 1(X)$ $\delta 3$ *carries* the level x-which is present in assertion X- just above the corresponding level in $\delta 2(X)$, and so on.
Equation (IIe) gives the rule of composition for maps $\delta t$.
Equation (IIf) states that any two assertions which have equal images under all maps $\delta t$, are equal.

Equation (IIg) states that the application of $\delta i$ to the negation of proposition X leads to the negation of proposition $\delta (X)$, if i+j = *n*-1.

The behavior of a genetic network can also be intuitively pictured by n table with *k* columns, corresponding to the genes of the net, and with rows corresponding to the moments which are counted backwards from the present moment p. The positions in the table are filled with 0's, l's and letters i,j, . . ., (n-1) which stand for levels in the activity of genes. Thus, <u>1</u> denotes the *i*-th gene maximal activity. For example, with *k* = 3, the table might be as in Table I.

**Table I. A table representation of the behavior of the particular genetic net**

| Time | A | B | C |
|------|---|---|---|
| P    | 0 | .1 | i |
| P-ε  | k | 0 | 1 |
| P-δ  | 1 | 0 | 1 |
| …    |   |   |   |

The 0 in the first row and the first column means that gene *A* is inactive at time *p*; the 1 in the first row and second column means that *C* is active on the *i*-th level of intensity, at the same moment. In order to characterized mutations of genetics networks one has to consider mappings on n-valued Lukasiewicz algebras. These lead, in turn, to categories of genetic networks that contain all such networks together with all of their possible transformations and mutations.

(D2) A mapping **f**: $L_1 \to L_2$ is called a ***morphism of Łukasiewicz algebras*** if it has the following properties:

(M1). $f(0)=0$, $f(1)=1$, $f*N = N*f$;

(M2). $F(X \cup Y) = f(X) \cup f(Y)$; $f(X \cap Y) = f(X) \cap f(Y)$, for any $X, Y \in L$;

(M3). $f*d = d*f$, for any $y = 0, 1, 2, \ldots, n-1$.

The totality of mutations of genetic nets is then represented by a subcategory of $Luk_n$ – the category of n-valued Łukasiewicz algebras and morphisms among these, as discussed next in *Section 3*. A special case of n-valued Łukasiewicz algebras is that of centered Łukasiewicz algebras, that is, these algebras in which there exist (n-2) elements $a_1, a_2, \ldots a_n \varepsilon$ : (called centers), such that:

$$\delta(a_j) = \begin{cases} 0, & \text{for } 1 \leq j \leq n-j \\ 1, & \text{for } n-j \leq i \leq n-2. \end{cases}$$

If the activity of genes would be of the "all or none" type then we would have to consider genetic nets as represented by Boolean algebra. A subcategory of $B_1$, the category of Boolean algebras, would then be represented by the totality of mutations of "all or none" type of genes. However, there exists equivalence between the category of centered Lukasiewicz algebras.

This equivalence is expressed by two adjoint functors $Luk_n \xrightarrow{C} Bl \xrightarrow{D} Luk_n$
with C being full and faithful (Georgescu and Vraciu, 1970). The above algebraic result shows that the particular case n=2 (that is "all or none" response) can be treated by means of centered Łukasiewicz algebras.

## 1. Categories and Topoi of Genetic Networks

Let us consider next categories of genetic networks that are collections of such networks and their functional transformations. These are in fact subcategories of $Luk_n$, the category of Łukasiewicz Logic Algebras and their connecting "morphisms". The totality of the genes present in a given organism—or a genome-can thus be represented as an object in the associated category of genetic networks of that organism. Let us denote this category of genetic networks by N, and call it the **genetic transformation category**. There exists a genetic network and its associated transformations in N that corresponds to the fertilized ovum form which the organism developed. This genetic net will be denoted by **0**, or **$G_0$**.

*Theorem 1.* **The Category N of Genetic Networks of any organism has a projective limit**.

*Proof.* To prove this theorem is to give an explicit construction of the genetic net which realizes the projective limit. If $G_1$, $G_2$,…,$G_i$ are distinct genetic nets, corresponding to different stages of development of a. certain organism, then let us define the cartesian product of the last ($l$-1) genetic nets $\prod G_j$ as the product of the underlying lattices $L_2$, $L_3$…, $L_p$. Correspondingly, we have now ($l$-1) tuples are formed with the sentences present in $L_2$, $L_3$,…$L_p$, as members.

The theorem is proven by the commutativity of the diagram for any $G_k$ and $G_m$ in the sequence $G_2$, $G_3$,…..$G_i$ such that m>k. The commutativity of this diagram is compatible with conditions (M1), (M2) and (M3) that define morphisms of lattices. Moreover,

$$G_i = \coprod_{i=0}^{1} G_i$$

and one also has that $G_i = 0$. Q.E.D.

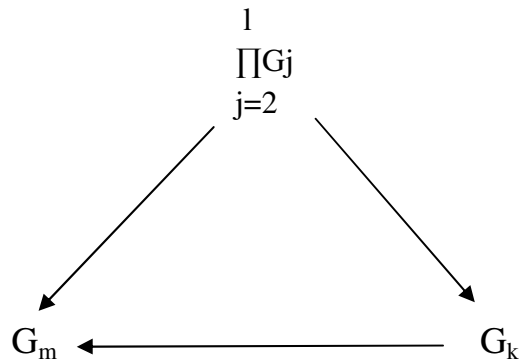

This result shows that the genetic network corresponding to a fertilized ovum is the projective limit of all subsequent genetic networks-corresponding to later stages of development of that organism.

Such an important algebraic property represents the 'potentialities for development of a fertilized ovum'.

*Theorem 2.* **Any family of Genetic Networks of N has a direct sum, and also a cokernel exists in** N.

The proof is immediate and stems from the categorical definitions of direct sum and cokernel (Mitchell,1965; and Baianu, 1970, 1977, in the context of organismic models). The above two theorems show a dominant feature of the category of genetic nets. The algebraic properties of N are similar to those exhibited by the category of all automata (sequential machines), and by its subcategory of (M, R)-systems, **MR** (for details see theorems 1 and 2, Baianu, 1973).

Furthermore, Theorems 1 and 2 hint at a more fundamental conjecture stating that: "There exist **adjoint functors** (Baianu,1970) between the category of genetic networks described here and the category of (M,R)-systems characterized previously (Theorems 1 and 2 of Baianu, 1977, and Baianu,1973, respectively); there are also certain Kan extensions of the (M,R)-systems category in the N, and $Luk_n$, categories that could be constructed explicitly for specific equivalent classes of (M,R)-systems and their underlying, adjoint genetic networks". Such Kan extensions may be restricted to the subcategory of centered Łukasiewicz Logic Algebras and their Boolean-compatible dynamic transformations of (M,R)-systems, with the latter as defined by Rosen (1971, 1973).

## *4. Realizability of Genetic Networks*.

The genes in a given network $G$ will be relabeled in this section by $g_1, g_2, g_3, \ldots g_N$. The *peripheral* genes of G are defined as the genes of G which are not influenced by the activity of other genes, and that in their turn do not influence more than one gene by their activity. Such genes have connectivities that are very similar to those present in random genetic networks, and could be presumably studied in Łukasiewicz Logic extensions of random genetic networks, rather than in strictly Boolean logic nets. The intermediate case of centered Łukasiewicz Algebra models of random genetic networks will thus provide a seamless link between various type of logic-based random networks, and also to Bayesian analysis of simpler organism genomes, such as that of yeast, and possibly *Archeas* also.

The assertion A(t;0) in (1) is called *the action* of gene $g_A$. The predicates which define the activities of genes comprise their *syntactical class*. As in the formalization inspired by McCullouch and Pitts, a *solution of G* will be a class of sentences of the form:

$S_t: A_{p+1}(z1). \equiv . Pr_i(A, B, \ldots, N_p; Z_n)$

with $Pr_i$ being a predicate expression which contains no free variable save $z_1$, and such that $S_t$ has one of the values of this Łukasiewicz n-Logic, except zero. The functor S is defined by the two following equalities:

$S(P)(t;k). \equiv . P(kx)_k . t = x$

$$S^2 Pr = S(S(Pr)), \ldots, S^k(Pr) = S(S(\ldots(S(Pr)))$$

$$\longleftrightarrow \text{ k-times}$$

Given a predicate expression $S^m(Pr1)(P_1,\ldots,Pp,z1)$, with **m** a natural number and **s** a constant sequence, then it is said to be *realizable* if there exists a genetic, or neural, network **G** and a series of activities such that :

$$A1(z1) \equiv Pr1(A1, A2, \ldots, z1, s_{a1})$$

has a non-zero logical value for $s_{a1} = A(0)$. Here the *realizing gene* will be denoted by $g_{p1}$.

Two laws concerning the activities of the genes, which are such that every S which is realizable for one of them is also realizable for the other, will be called **equivalent**. Equivalent genes may have additional algebraic structures in terms of **topological grupoids** (that is, categories consisting of topological space isomorphisms; Ehresmann, 1956; Brown, 1975) and subcategories of Lukn that contain such topological grupoids of equivalent genes, **TopGd.**

A genetic network will be called **cyclic** if each gene of the net is arranged in a functional chain with the same beginning and end. In a cyclic **net** each gene acts on its next neighbor and is influenced by its precedent neighbor. If a set of genes $g_1, g_2, g_3, \ldots, g_p$ of the genetic net **G** is such that its removal from G leaves G without cycles, and if no proper subset has this property, then the set is called **cyclic**. The cardinality of this set is an index on the complexity of its behavior. It will be seen later that this index does not uniquely determine the complexity of behavior of a genetic network. Furthermore, such cyclic subnetworks of the genome may have additional algebraic structure that can be characterized by a certain type of algebraic groups that will be called genetic groups, and will be forming a Category of Genetic Groups, **GrG**, with group transformations as group morphisms. **GrG** is obviously a subcategory of **N**, the genetic network transformation category, or the category of time-dependent genomes. In its turn, the category N is a subcategory of the higher order Cell Interactome category, **IntC**, that includes all signaling pathways coupled to the genetic networks, as well as their dynamic transformations and other metabolic components and processes essential to cell survival, growth, development, division and differentiation.

There is, therefore, in terms of the organizational hierarchy and complexity indices of the various categories of networks
the following partial, and strict, ordering:

**Automata Semigroup Category (ASG) $\leq$ MR $\leq$ CtrLukn < GrG < TopGd < IntC <Lukn**

This sequence of network structure models forms a finite, organizational semi-lattice of subcategories of network models in **Lukn**. Their classification can be effectively carried out by selecting the Łukasiewicz Logic Algebras as the *subobject classifier* in a **Łukasiewicz Logic Algebras Topos** (Baianu et al, 2004) that includes the cartesian closed category (Baianu,1973) of all networks that has limits and colimits. A particularly interesting example is that of the **TopGd** category that will contribute certain associated sheaves of genetic networks with striking, 'emerging' properties such as 'genetic memory' that perhaps reflects underlying holonomic *quantum genetic proceeses,* as well as

*related quantum automata* **reversibility** properties, such as enzyme reaqctions and/or *relational oscillations* in genetic networks during cell cycling (Baianu, 1971), neoplastic transformations of cells and carcinogenesis (Baianu, 1971,1977).

(D3) An *n-valued propositional expression* (NTPE) designates a **temporal propositional function** (TPF) and is defined by the following recursion:

(NT1). A *1p1[z]* is an NTPE if *P1* is a predicate variable with n-possible logical values;

(NT2). If S1 and S2 are NTPE containing the same free individual variable, so are S1 $\wedge$ S2, S1 $\vee$ S2, S1.S2, and S1~S2.

One can easily prove the following theorems.

**Theorem 3.** Every genetic, net of order zero can be solved in terms of *n-valued temporal propositional expressions* (NTPE).

**Theorem 4**. *Every NTPE is realizable in terms of a genetic net of zero-th order*.

**Theorem** 5. **Any** *complex sentences* **S1** *(built up in* **any** *manner out of elementary sentences of the form* **p(z1-zz),** *(where zz is any numeral), by means of negation, conjunction, implication and logical equivalence),* **is** *an* **NTPE**.

$S_i$ acquires zero value only when all its constituents p(z1-zz0 have all the zero logical value ( "false"). Let us recall that if two or more genes influence the activity of the same gene, then the influenced genes are said to be **alterable**. One readily obtains the following theorem concerning alterable genes:

**Theorem 6.** *Alterable genes can be replaced by cycles*..

For cyclic genetic nets of order **p** one can generalize the construction method proposed by McCullouch and Pitts. However, there will be no different sentences formed out of the pN1 by joining to the conjunction of some set of the conjunctions of the "negated" forms of each level of the rest. Consequently, the logical expression which is a solution of G, will have the form:

$(z4)(z1)zzp;Pri(z_i,z4) \equiv (\exists f)(z_{i+1})[z(i+3)-1]f(z(i+1))$ ,

with i =1,2,. .., (n- 3). zzp, res (r, s) is the residue of r mod s and zzp=ip .

In our case the realizability of a set of Si is not simple as it was in the case of Boolean logic, neural nets. Now, it involves **n** simultaneous conditions for the **n** distinct logical values, instead of just the two values from Boolean logic. As a consequence, it is possible that certain genetic networks will be able to 'take into account' the future of their peripheral genes in their switching sequence and levels of activities, thus effectively anticipating sudden threats to the cell survival, and also exhibiting multiple adaptation behaviors in response to exposure to several damaging chemicals or mutagens, antibiotics, etc. Thus, another index of complexity of behavior of genetic networks is the number of *future* peripheral genes which are taken into account by a specific realization of a network. In contrast to a feedback system, this will be called a *feedforward* system. Furthermore, the fact that the number of active genes, or simply the number or genes, is not constant in an organism during its development, but increases until maturity is reached, makes it difficult to apply directly the 'purely' logical formalization introduced in this section. However, the categorical and Łukasiewicz-Logic Topos formalization that was introduced in Section 2 can now be readily applied to developmental processes and effectively solves such realizability problems through effective categorical construction methods such as presheaves, sheaves, higher dimensional algebras, limits, colimits, adjoint functors and Kan extensions.

## 5. *A Specific Example of Genetic Network Module present in Cancer Cells*

Figure 1 of Baianu & Prisecaru (2004) illustrates a specific case of a genetic network module representing Cancer Cycling, Cell Division Control and p53 Functions. Baianu and Prisecaru, 2004*, arXiv.q-bio/0406046)* .that has been reported in several types of the most common cancers. Further details are presented in the accompanying paper The analysis of such a module can proceed both through qualitative dynamics tools and programes (such as GNA) or through the algebraic-theoretical and Łukasiewicz-Logic analysis in Topoi that was introduced in the previous section.

## *6. Discussion and Conclusions*

A significant application of related to Boolean Logic was the calculus of predicates which was applied by Nicolas Rashevsky (1965) to more general situations in relational biology and organismic set theory. Lőfgren (1968) introduced also a non-Boolean logical approach to the problem of self-reproduction. The characterization of genetic activities in terms of Łukasiewicz Logic Algebras that was here presented has only certain broad similarities to the well known method of McCulloch and Pitts (1943). There are major differences arising in genetic networks both from the fact that the genes are considered to act in a step-wise manner, as well as from the coupling of the genetic network to the cell interactomics through intracellular signaling pathways. The "all-or-none" type of activity often considered in connection with genes results as a particular case of the generalized description for $n = 2$ in centered Łukasiewicz logic algebras. The new concept of a Łukasiewicz Topos expands the applications range of such models of genetic activities to whole genome, cell interactomics, neoplastic transformations and morphogenetic or evolutionary processes.

The approach of genetic activities from the standpoint of Łukasiewicz Logic algebras categories and Topoi leads to the conclusion that the use of n-valued logics for the description of genetic activities allows for the emergence of new algebraic and transformation properties that are in agreement with several lines of experimental evidence (such as adaptability of genetic nets and feedforward, or anticipatory, processes), including evolutionary biology observations, as well as a wide array of cell genomic and interactomic data for the simpler organisms, such as yeast and a nematode (*C. elegans*) species. In principle, and hopefully soon, in practice, such categorical- and Topos- based applications to *cell genomes and interactomes* will not be limited to the simpler organisms but will also include higher organisms such as *Homo sapiens*.

Nonlinear dynamics of non-random genetic and cell networks can be thus formulated explicitly through categorical constructions enabled by Łukasiewicz Logic algebras that are in principle computable through symbolic programming on existing high performance workstations and supercomputers even for modeling networks composed of huge numbers of interacting 'biomolecular' species (Baianu et al., 2004). Strategies for meaningful measurements and observations in real, complex biological systems (Baianu et al., 2004 a), such as individual human organisms, may thus be combined with genomic and proteomic testing on individuals and may very well lead to optimized, individualized therapies for life-threatening diseases such as cancer and cardiovascular diseases.

On the other hand, one has to consider the fact that the problem of compatibility or solvability of complex models is further complicated by the presence of n-valued logics. The categorical notion of **representable functor** would correspond to the computability concept for genetic nets. This strongly indicates that the genetic nets are not generally equivalent to Turing machines as the neural nets are. However, the results of *Section 3* show that only those genetic networks that are characterized completely by centered Łukasiewicz algebras may possess equivalent Turing machines. In the case of our n-state model the realizability of a set of Si is not simple as it was in the case of *Boolean* logic, neural nets. Now, it involves **n** simultaneous conditions for the **n** distinct logical values, instead of just the two values from *Boolean* logic. As a consequence, it is possible that certain genetic networks will be able to 'take into account' the future of their peripheral genes in their switching sequence and levels of activities, thus effectively anticipating sudden threats to the cell survival, and also exhibiting multiple adaptation behaviors in response to exposure to several damaging chemicals or mutagens, antibiotics, etc. Thus, another index of complexity of behavior of genetic networks is the number of *future* peripheral genes which are taken into account by a specific realization of a network. In contrast to a feedback system, this will be called a *feedforward* system. Furthermore, the fact that the number of active genes, or simply the number or genes, is not constant in an organism during its development, but increases until maturity is reached, makes it difficult to apply directly the 'purely' logical formalization introduced in this section. However, the categorical and Łukasiewicz-Logic Topos formalization that was introduced in Section 2 can now be readily applied to developmental processes and effectively solves such realizability problems through effective categorical construction methods such as presheaves, sheaves, higher dimensional algebras, limits, colimits, adjoint functors and Kan extensions (Baianu et al, 2004, in preparation ). The "all-or-none" type of activity often considered in connection with genes results as a particular case of the generalized description for $n = 2$ in centered Łukasiewicz logic algebras.

The new concept of a Łukasiewicz-Topos expands the applications range of such models of genetic activities to whole genome, cell interactomics, neoplastic transformations and morphogenetic or evolutionary processes. The approach of genetic activities from the standpoint of Łukasiewicz Logic algebras categories and Topoi leads to the conclusion that the use of n-valued logics for the description of genetic activities allows for the emergence of new algebraic and transformation properties that are in agreement with several lines of experimental evidence (such as, adaptability of genetic nets and feedforward, or anticipatory, processes), including evolutionary biology observations, as well as a wide array of cell genomic and interactomic data for the simpler organisms, such as yeast and a nematode (*C. elegans*) species. In principle, and hopefully soon, in practice, such categorical- and Topos- based applications to *cell genomes and interactomes* will not

be limited to the simpler organisms but will also include higher organisms such as *Homo sapiens.*

Nonlinear dynamics of non-random genetic and cell networks can be thus formulated explicitely through categorical constructions enabled by Łukasiewicz Logic algebras that are in principle computable through symbolic programming on existing high performance workstations and supercomputers even for modeling networks composed of huge numbers of interacting 'biomolecular' species (Prisecaru and Baianu, 2004, arXiv. Q-bio.0406046 *Preprint*). Strategies for meaningful measurements and observations in real, complex biological systems (Baianu et al., 2004 ), such as individual human organisms, may thus be combined with genomic and proteomic testing on individuals and may very well lead to optimized, individualized therapies for life-threatening diseases such as cancer and cardiovascular diseases.

On the other hand, one has to consider the fact that the problem of compatibility or solvability of complex models is further complicated by the presence of n-valued logics. The categorical notion of representable functor would correspond to the computability concept for genetic nets. This strongly indicates that the genetic nets are not generally equivalent to Turing machines as the neural nets are. However, the results of *Section 3* show that only those genetic networks that are characterized completely by centered Łukasiewicz algebras may possess equivalent Turing machines.

The formalization introduced in Sections 2 and 3 in terms of categories, functors, higher dimensional algebra and Łukasiewicz Topos, (and probably also intuitionistic,Heyting Logic Topoi), allows additional, important results to be obtained which will be presented in a subsequent paper.